\documentclass[prd,amsmath,amssymb,superscriptaddress,preprintnumbers,twocolumn,10pt]{revtex4}
\pdfoutput=1
\usepackage{graphicx}
\usepackage{dcolumn}
\usepackage{bm}
\usepackage{amssymb}
\usepackage{latexsym}
\usepackage{booktabs}
\usepackage{amsmath}
\usepackage{multirow}
\usepackage{url}
\usepackage{float}
\usepackage[colorlinks=true, linkcolor=red, citecolor=blue]{hyperref}

\newcommand{\ve}{\varepsilon}
\usepackage[normalem]{ulem}
\usepackage{color}
\usepackage{array}
\usepackage{enumerate}

\begin{document}

\title{Null test for cosmic curvature using Gaussian process}

\author{Peng-Ju Wu}
\affiliation{Key Laboratory of Cosmology and Astrophysics (Liaoning Province) \& Department of Physics, College of Sciences, Northeastern University, Shenyang 110819, China}

\author{Jing-Zhao Qi}
\affiliation{Key Laboratory of Cosmology and Astrophysics (Liaoning Province) \& Department of Physics, College of Sciences, Northeastern University, Shenyang 110819, China}

\author{Xin Zhang}\thanks{Corresponding author.\\zhangxin@mail.neu.edu.cn}
\affiliation{Key Laboratory of Cosmology and Astrophysics (Liaoning Province) \& Department of Physics, College of Sciences, Northeastern University, Shenyang 110819, China}
\affiliation{Key Laboratory of Data Analytics and Optimization for Smart Industry (Ministry of Education), Northeastern University, Shenyang 110819, China}
\affiliation{National Frontiers Science Center for Industrial Intelligence and Systems Optimization, Northeastern University, Shenyang 110819, China}

\begin{abstract}
The cosmic curvature $\Omega_{K,0}$, which determines the spatial geometry of the universe, is an important parameter in modern cosmology. Any deviation from $\Omega_{K,0}=0$ would have a profound impact on primordial inflation paradigm and fundamental physics. In this work, we adopt a {cosmological} model-independent method to test whether $\Omega_{K,0}$ deviates from zero. We use the Gaussian process to reconstruct the reduced Hubble parameter $E(z)$ and the derivative of distance $D'(z)$ from observational data, and then determine $\Omega_{K,0}$ with a null test relation. The cosmic chronometer (CC) Hubble data, baryon acoustic oscillation (BAO) Hubble data, and supernovae Pantheon sample are considered. Our result is consistent with a spatially flat universe within the domain of reconstruction $0<z<2.3$, at the $1\sigma$ confidence level. In the redshift interval $0<z<1$, the result favors a flat universe, while at $z>1$, it tends to favor a closed universe. In this sense, there is still a possibility for a closed universe. We also carry out the null test of the cosmic curvature at $0<z<4.5$ using the simulated gravitational wave standard sirens, CC+BAO and redshift drift Hubble data. The result shows that in the future, with the synergy of multiple high-quality observations, we can tightly constrain the spatial geometry or exclude the flat universe.
\end{abstract}

\maketitle
\section{Introduction}\label{sec1}
The cosmic curvature $\Omega_{K,0}$ is an important parameter that is related to many fundamental problems in modern cosmology. Knowing whether the universe is spatially open ($\Omega_{K,0}>0$), flat ($\Omega_{K,0}=0$), or closed ($\Omega_{K,0}<0$) is crucial for us to understand its evolution and the property of dark energy. A flat universe is strongly favored by some cosmological observations. For instance, the Planck 2018 cosmic microwave background (CMB) observations combined with the baryon acoustic oscillations (BAO) measurements give $\Omega_{K,0}=0.0007\pm0.0019$, suggesting that our universe is flat to a $1\sigma$ error of $2\times 10^{-3}$ \citep{Aghanim:2018eyx}. However, it was found that the Planck TT,TE,EE+lowE power spectra data alone favors a slightly closed universe, $\Omega_{K,0}=-0.044_{-0.015}^{+0.018}$ \citep{Aghanim:2018eyx, Park:2017xbl, Handley:2019tkm, DiValentino:2019qzk}. This deviation from a flat universe is interpreted as the undetected systematics, statistical fluctuation, or new physics beyond the $\Lambda$ cold dark matter ($\Lambda$CDM) model. Efstathiou and Gratton \citep{Efstathiou:2020wem} recently revisited the issue and claimed that the Planck data are still consistent with a flat universe. Whether this crisis really exists is still under debate.

It should be pointed out that most of the curvature parameter estimations assume a specific cosmological model. However, there is a strong degeneracy between the curvature parameter and the dark energy equation of state $w(z)$, so it is difficult to constrain them simultaneously, which hinders our understanding of dark energy. Therefore, it is necessary to measure the cosmic curvature in a cosmological model-independent way. For this purpose, many novel and feasible methods have been proposed; see e.g., Refs.~\citep{Bernstein:2005en, Shafieloo:2009hi, Li:2014yza, Li:2016wjm, Li:2019bbg, Sapone:2014nna, Rasanen:2014mca, Pan-STARRS1:2017jku, Yu:2016gmd, LHuillier:2016mtc, Liao:2019hfl, Rana:2016gha, Wang:2019yob, Wei:2016xti, Xia:2016dgk, Denissenya:2018zcv, Wei:2018cov, Witzemann:2017lhi, Collett:2019hrr, Qi:2018aio, Wei:2020suh, Zhou:2019vou, Dhawan:2021mel, Jesus:2019jvk, Vagnozzi:2020dfn, Zhao:2021jeb, Wei:2022rcb, Wei:2022plg, Koksbang:2022ari, Liu:2022lqw, Zhang:2022lta}.

In Ref.~\citep{Cai:2015pia}, Cai et al. proposed a model-independent method to test whether the cosmic curvature deviates from zero. They adopted the observational data to reconstruct the reduced Hubble parameter $E(z)$ and distance-redshift relation $[D(z), D'(z)]$, and then combined the reconstructions to perform the null test of $\Omega_{K,0}$. {In their analysis, the $H(z)$ measurements from cosmic chronometer (CC) and BAO observations as well as the Union2.1 type Ia supernovae (SNe Ia) sample are considered, and the result favors a flat universe. Due to the increase of observational data, Yang and Gong \citep{Yang:2020bpv} reperformed the null test using the CC $H(z)$ data and SNe Ia Pantheon compilation, and the result is still consistent with a flat universe. In Ref.~\citep{Yang:2020bpv}, the BAO $H(z)$ data are not considered. It should be pointed out that the CC data may not constitute a reliable source of information due to some concerns \citep{Kjerrgren:2021zuo}. In contrast, the $H(z)$ data from BAO observations are much more accurate and reliable. In the present work, we consider both the CC and BAO $H(z)$ measurements, i.e., we shall use the latest CC+BAO $H(z)$ data and SNe Ia Pantheon sample to test the spatial flatness of the universe.}

{Furthermore, we also explore what role the future gravitational wave (GW) standard sirens and CC+BAO+redshift drift (RD) observations will play in the null test of $\Omega_{K,0}$.} GWs can serve as standard sirens, since the GW waveform carries the information of the luminosity distance $D_L$ to source \citep{Schutz:1986gp,Holz:2005df,Zhang:2019ylr}. If the source's redshift can be determined, for example, by identifying the electromagnetic counterpart of the GW event, we can then establish the $D_L$-redshift relation. We simulate the GW $D_L(z)$ data based on the planned space-based GW detector, DECihertz Interferometer Gravitational wave Observatory (DECIGO)~\citep{Kawamura:2020pcg}. {In the coming decades, with the advent of some powerful optical and radio telescopes, such as the Euclid~\citep{Amendola:2016saw}, Subaru Prime Focus Spectrograph (PFS)~\citep{PFSTeam:2012fqu}, Dark Energy Spectroscopic Instrument (DESI)~\cite{DESI:2016fyo,DESI:2016igz}, and Square Kilometre Array (SKA)~\cite{SKA:2018ckk}, we can better measure the Hubble parameter using the CC and BAO methods. We simulate the CC+BAO $H(z)$ data in the redshift interval $0<z<2.5$ according to the observational data.} In the future, another promising way to measure $H(z)$ is the RD method \citep{Loeb:1998bu}. We simulate the high-$z$ Hubble data based on hypothetical RD observations of the upcoming European Extremely Large Telescope (E-ELT). We shall use the simulated GW and CC+BAO+RD data to perform the null test of $\Omega_{K,0}$ and compare the result with that using the current CC+BAO and SNe Ia data.

In this work, we adopt a machine learning method, the Gaussian process (GP), to reconstruct the cosmological functions. It has been widely used in cosmological researches; see e.g., Refs.~\citep{Holsclaw:2010sk, Shafieloo:2012ht, Seikel:2012uu, Yahya:2013xma, Yang:2015tzc, Cai:2015pia, Wang:2017lri,Elizalde:2018dvw,Elizalde:2018ahd, Liao:2019qoc, Cai:2019bdh, Mukherjee:2020ytg, Mukherjee:2020vkx, Mehrabi:2021cob, Aljaf:2020eqh, Vazirnia:2021xuu, Mukherjee:2021kcu, Mukherjee:2021ggf, Ruiz-Zapatero:2022zpx, Hwang:2022hla, Mukherjee:2022ujw,Elizalde:2022rss,Elizalde:2022oej}. The GP method allows one to reconstruct a function and its derivative from data without assuming any particular parametrization, so it is suitable for our purpose. {The artificial neural network (ANN) has recently emerged as a promising tool for reconstructing functions \citep{Wang:2019vxv, Wang:2020dbt,Liu:2021fka,Qi:2023oxv}, however, as far as we know, ANN has difficulties in reconstructing the derivative of a function. For this reason, our research is based on the GP analysis.}

The remainder of this paper is organized as follows. We briefly describe the methodology in Sec.~\ref{sec2}. Sec.~\ref{sec3} contains the data we adopted. We present the results and make some discussions in Sec.~\ref{sec4}. Finally, we give our conclusions in Sec.~\ref{sec5}.

\section{Methodology}\label{sec2}
In the homogeneous and isotropic universe, the FLRW metric is applied to describe its spacetime:
\begin{align}
d s^{2}=-c^2d t^{2}+a^{2}(t)\left[\frac{d r^{2}}{1-K r^{2}}+r^{2}\left(d \theta^{2}+\sin ^{2} \theta d \phi^{2}\right)\right],
\end{align}
where $c$ is the speed of light, $a$ is the scale factor, and $K$ is a constant that is related to the cosmic curvature by $\Omega_{K} \equiv -Kc^2 / (aH)^2$, with $H$ being the Hubble parameter. We use $\Omega_{K,0}$ to represent the present value of $\Omega_{K}$, and then $\Omega_{K,0}>0$, $\Omega_{K,0}=0$ and $\Omega_{K,0}<0$ correspond to the open, flat and closed universe, respectively. The luminosity distance can be expressed as
\begin{align}
\label{DL}
D_{L}(z)=\frac{c(1+z)}{H_{0}\sqrt{\left|\Omega_{K,0}\right|}} \operatorname{sinn}\left[\sqrt{\left|\Omega_{K,0}\right|} \int_{0}^{z} \frac{d z^{\prime}}{E\left(z^{\prime}\right)}\right],
\end{align}
where
\begin{align}
\operatorname{sinn}(x)=\left\{\begin{array}{ll}
\sinh (x) , & \Omega_{K,0}>0 ,  \\
x, & \Omega_{K,0}=0 , \\
\sin (x), & \Omega_{K,0}<0 ,
\end{array}\right.
\end{align}
and $E(z)\equiv H(z)/H_0$ is the reduced Hubble parameter.

Differentiating Eq.~(\ref{DL}), we have \citep{Clarkson:2007bc,Clarkson:2007pz}
\begin{align}
\label{Hz_DL}
\Omega_{K,0}=\frac{E^{2}(z) D^{\prime 2}(z)-1}{D^{2}(z)},
\end{align}
where $D(z)=(H_0/c)D_{L}(z)(1+z)^{-1} $ is the dimensionless comoving distance. Obviously, the curvature parameter can be directly determined by using the Hubble parameter and luminosity distance according to Eq.~(\ref{Hz_DL}). Thus, we can perform the null test of $\Omega_{K,0}$. Note that $D(0)=0$ will bring a singularity at $z=0$. For simplicity, we transform Eq.~(\ref{Hz_DL}) to
\begin{align}
\label{refine}
\frac{\Omega_{K,0} D^{2}(z)}{E(z) D^{\prime}(z)+1}=E(z) D^{\prime}(z)-1.
\end{align}
We can see that the left-hand side of Eq.~(\ref{refine}) is non-zero when $z \neq 0$ if $\Omega_{K,0}$ is nonvanishing. Therefore, the null test of $\Omega_{K,0}$ is equivalent to the null test of the left-hand side of Eq.~(\ref{refine}). Following Cai et al. \citep{Cai:2015pia}, we define
\begin{align}
\label{xxx}
\mathcal{O}_{K}(z) \equiv \frac{\Omega_{K,0} D^{2}(z)}{E(z) D^{\prime}(z)+1}.
\end{align}
For a spatially flat universe,
\begin{align}
\label{test}
\mathcal{O}_{K}(z)=E(z) D^{\prime}(z)-1=0
\end{align}
is always true at any redshift, so the deviation from it will imply a nonvanishing cosmic curvature. To carry out the null test of Eq.~(\ref{test}), we need to reconstruct the functions of $E(z)$, $D(z)$, and $D'(z)$ using the observational data. In this work, we adopt the GP method to reconstruct the cosmological functions without assuming a specific cosmological model. Therefore, a cosmological model-independent test of whether $\Omega_{K,0}$ deviates from zero is performed in this work; note here that we do not assume a specific dark energy model, but of course an FLRW model of the homogeneous and isotropic universe is adopted.

{GP is a non-parametric smoothing method for reconstructing functions \citep{Holsclaw:2010sk, Shafieloo:2012ht, Seikel:2012uu, Yahya:2013xma, Yang:2015tzc, Cai:2015pia, Wang:2017lri,Elizalde:2018dvw,Elizalde:2018ahd, Liao:2019qoc, Cai:2019bdh, Mukherjee:2020ytg, Mukherjee:2020vkx, Mehrabi:2021cob, Aljaf:2020eqh, Vazirnia:2021xuu, Mukherjee:2021kcu, Mukherjee:2021ggf, Ruiz-Zapatero:2022zpx, Hwang:2022hla, Mukherjee:2022ujw,Elizalde:2022rss,Elizalde:2022oej}, which assumes that at each point $z$, the reconstructed function $f(z)$ is a Gaussian distribution. Furthermore, the functions at different points are related by a covariance function.} We can model a data set using a GP as
\begin{align}
f(z) \sim \mathcal{G} \mathcal{P} (\mu(z), k(z, \tilde{z})),
\end{align}
where $\mu(z)$ is the mean function which provides the mean of random variables at each observational point, and $k(z, \tilde{z})$ is the covariance function which correlates the values of different $f(z)$ at data points $z$ and $\tilde{z}$ separated by $|z-\tilde{z}|$ distance units.

There is a wide range of possible covariance functions, and the choice of covariance function actually affects the reconstruction to some extent \citep{Seikel:2012uu}. Here, we consider the commonly used squared exponential covariance function,
\begin{align}
k(z, \tilde{z})=\sigma_{f}^{2} \exp \left[-\frac{(z-\tilde{z})^{2}}{2 \ell^{2}}\right],
\end{align}
where $\sigma_{f}$ denotes the overall amplitude of the oscillations around the mean and $\ell$ gives a measure of the correlation length between the GP nodes. Both $\sigma_{f}$ and $\ell$ are hyperparameters, which will be optimized by GP with the observational data. Given the GP for $f(z)$, the GP for the first derivative is consequently given by
\begin{align}
f^{\prime}(z) \sim \mathcal{G} \mathcal{P}\left(\mu^{\prime}(z), \frac{\partial^{2} k(z, \tilde{z})}{\partial z \partial \tilde{z}}\right).
\end{align}
Therefore, it is convenient to calculate the derivative of reconstructed function. In this work, we use the publicly available {\tt GaPP} code to implement our analysis \citep{Seikel:2012uu}.

{The specific calculation process of {\tt GaPP} is as follows. For a set of input points, $\boldsymbol{Z}=\left\{z_{i}\right\}$, the covariance matrix $K(\boldsymbol{Z}, \boldsymbol{Z})$ is calculated by $[K(\boldsymbol{Z}, \boldsymbol{Z})]_{i j}=k\left(z_{i}, z_{j}\right)$. Even in the absence of observations, we can generate a random function $f(z)$ from GP, i.e., we can generate a vector $\boldsymbol{f^*}$ of function values at $\boldsymbol{Z^*}=\left\{z_{i}^*\right\}$ with $f_{i}^{*}=f\left(z_{i}^{*}\right)$:
\begin{align}
\label{priori}
\boldsymbol{f}^{*} \sim \mathcal{G P}\left(\boldsymbol{\mu}^{*}, K\left(\boldsymbol{Z}^{*}, \boldsymbol{Z}^{*}\right)\right),
\end{align}
where $\boldsymbol{\mu}^{*}$ is a prior of the mean of $\boldsymbol{f}^{*}$, which is set to zero in this paper. For observations, $\left\{(z_{i}, y_{i}, \sigma_i)|_{i=1, \ldots, N}\right\}$, one can also use a GP to describe them. We stress that $y_{i}$ is assumed to be scattered around the underlying function, i.e., $y_i = f(z_i)+\epsilon_{i}$, where Gaussian noise $\epsilon_{i}$ with variance $\sigma_i^2$ is assumed. Therefore, we need to add the variance to the covariance matrix,
\begin{align}
\label{data}
\boldsymbol{y} \sim \mathcal{G P}(\boldsymbol{\mu}, K(\boldsymbol{Z}, \boldsymbol{Z})+C),
\end{align}
where $\boldsymbol{y}$ is the vector of $y_i$ values and $C$ is the covariance matrix of the data. For uncorrelated data, we use $C=\text{diag}(\sigma_{i}^{2})$.}

{The above two GPs for $\boldsymbol{f}^{*}$ and $\boldsymbol{y}$ can be combined in the joint distribution:
\begin{align}
\left[\begin{array}{c}
\boldsymbol{y} \\
\boldsymbol{f}^{*}
\end{array}\right] \sim \mathcal{G P}\left(\left[\begin{array}{c}
\boldsymbol{\mu} \\
\boldsymbol{\mu}^{*}
\end{array}\right],\left[\begin{array}{cc}
K(\boldsymbol{Z}, \boldsymbol{Z})+C & K\left(\boldsymbol{Z}, \boldsymbol{Z}^{*}\right) \\
K\left(\boldsymbol{Z}^{*}, \boldsymbol{Z}\right) & K\left(\boldsymbol{Z}^{*}, \boldsymbol{Z}^{*}\right)
\end{array}\right]\right).
\end{align}
Here $\boldsymbol{y}$ is known from observations. To reconstruct $\boldsymbol{f}^{*}$, one can consider the conditional distribution,
\begin{align}
\label{posterior}
\boldsymbol{f}^{*} \mid \boldsymbol{Z}^{*}, \boldsymbol{Z}, \boldsymbol{y} \sim \mathcal{G P}\left(\overline{\boldsymbol{f}^{*}}, \operatorname{cov}\left(\boldsymbol{f}^{*}\right)\right),
\end{align}
where
\begin{align}
\label{mean}
\overline{\boldsymbol{f}^{*}}=\boldsymbol{\mu}^{*}+K\left(\boldsymbol{Z}^{*}, \boldsymbol{Z}\right)[K(\boldsymbol{Z}, \boldsymbol{Z})+C]^{-1}(\boldsymbol{y}-\boldsymbol{\mu})
\end{align}
and
\begin{align}
\label{cov}
\operatorname{cov}\left(\boldsymbol{f}^{*}\right)&= K\left(\boldsymbol{Z}^{*}, \boldsymbol{Z}^{*}\right) \nonumber \\
&-K\left(\boldsymbol{Z}^{*}, \boldsymbol{Z}\right)[K(\boldsymbol{Z}, \boldsymbol{Z})+C]^{-1} K\left(\boldsymbol{Z}, \boldsymbol{Z}^{*}\right)
\end{align}
are the mean and covariance of $\boldsymbol{f}^{*}$, respectively. Eq.~(\ref{posterior}) is the posterior distribution of Eqs.~(\ref{priori}) and (\ref{data}). To reconstruct $\boldsymbol{f}^{*}$ using the above equations, we need to know the hyperparameters $\sigma_{f}$ and $\ell$, which can be determined by maximizing the logarithm marginal likelihood,
\begin{align}
\label{loglike}
\ln \mathcal{L}=& \ln p\left(\boldsymbol{y} \mid \boldsymbol{Z}, \sigma_{f}, \ell\right) \nonumber \\
=&-\frac{1}{2}(\boldsymbol{y}-\boldsymbol{\mu})^{T}[K(\boldsymbol{Z}, \boldsymbol{Z})+C]^{-1}(\boldsymbol{y}-\boldsymbol{\mu})\nonumber \\
&-\frac{1}{2} \ln |K(\boldsymbol{Z}, \boldsymbol{Z})+C|-\frac{N}{2} \ln 2 \pi.
\end{align}
{The hyperparameters will be fixed after being optimized through Eq.~(\ref{loglike}), then the reconstructed function $\boldsymbol{f}^{*}$ at the chosen points $\boldsymbol{Z}^{*}$ can be calculated from Eqs.~(\ref{mean}) and (\ref{cov}). Therefore, GP is model-independent and without free parameters.}

\section{Data}\label{sec3}
Here we present the data used for reconstructions. We do not consider to use the CMB data, because (i) we do not use observations to constrain a specific model, (ii) the reconstructions cannot be performed up to the early universe (e.g., the last scattering), and (iii) the inconsistencies in the measurements of the early- and late-universe observations (such as the well-known ``Hubble tension'') should also be considered. Thus, we only use the late-universe observations in this work.
In the first two subsections \ref{sec3.1} and \ref{sec3.2}, we present the CC+BAO and SNe Ia real data, and in the last two subsections \ref{sec3.3} and \ref{sec3.4}, we present the GW and CC+BAO+RD mock data.

\subsection{CC + BAO}\label{sec3.1}
\begin{table}
\renewcommand\arraystretch{1}
\caption{32 $H(z)$ measurements (in units of $\rm km\,s^{-1}\,Mpc^{-1}$) obtained with the CC method.}
\label{CCHz}
\centering
\begin{tabular}{l p{1.8cm}<{\centering} p{1.8cm}<{\centering} p{1.8cm}<{\centering} p{1.8cm}<{\centering}}
\bottomrule[1pt]
& Redshift $z$ & $H(z)$        & $\sigma_{H(z)}$       & Reference                       \\
\bottomrule[1pt]
& 0.07         & 69            & 19.6                  & \citep{Zhang:2012mp}             \\
& 0.09         & 69            & 12                    & \citep{Simon:2004tf}             \\
& 0.12         & 68.6          & 26.2                  & \citep{Zhang:2012mp}             \\
& 0.17         & 83            & 8                     & \citep{Simon:2004tf}             \\
& 0.179        & 75            & 4                     & \citep{Moresco:2012jh}           \\
& 0.199        & 75            & 5                     & \citep{Moresco:2012jh}           \\
& 0.2          & 72.9          & 29.6                  & \citep{Zhang:2012mp}             \\
& 0.27         & 77            & 14                    & \citep{Simon:2004tf}             \\
& 0.28         & 88.8          & 36.6                  & \citep{Zhang:2012mp}             \\
& 0.352        & 83            & 14                    & \citep{Moresco:2012jh}           \\
& 0.38         & 83            & 13.5                  & \citep{Moresco:2016mzx}          \\
& 0.4          & 95            & 17                    & \citep{Simon:2004tf}             \\
& 0.4004       & 77            & 10.2                  & \citep{Moresco:2016mzx}          \\
& 0.425        & 87.1          & 11.2                  & \citep{Moresco:2016mzx}          \\
& 0.445        & 92.8          & 12.9                  & \citep{Moresco:2016mzx}          \\
& 0.47         & 89            & 49.6                  & \citep{Ratsimbazafy:2017vga}     \\
& 0.4783       & 80.9          & 9                     & \citep{Moresco:2016mzx}          \\
& 0.48         & 97            & 62                    & \citep{Stern:2009ep}             \\
& 0.593        & 104           & 13                    & \citep{Moresco:2012jh}           \\
& 0.68         & 92            & 8                     & \citep{Moresco:2012jh}           \\
& 0.75         & 98.8          & 33.6                  & \citep{Borghi:2021rft}           \\
& 0.781        & 105           & 12                    & \citep{Moresco:2012jh}           \\
& 0.875        & 125           & 17                    & \citep{Moresco:2012jh}           \\
& 0.88         & 90            & 40                    & \citep{Stern:2009ep}             \\
& 0.9          & 117           & 23                    & \citep{Simon:2004tf}             \\
& 1.037        & 154           & 20                    & \citep{Moresco:2012jh}           \\
& 1.3          & 168           & 17                    & \citep{Simon:2004tf}             \\
& 1.363        & 160           & 33.6                  & \citep{Moresco:2015cya}          \\
& 1.43         & 177           & 18                    & \citep{Simon:2004tf}             \\
& 1.53         & 140           & 14                    & \citep{Simon:2004tf}             \\
& 1.75         & 202           & 40                    & \citep{Simon:2004tf}             \\
& 1.965        & 186.5         & 50.4                  & \citep{Moresco:2015cya}          \\
\bottomrule[1pt]
\end{tabular}
\end{table}

\begin{table}
\renewcommand\arraystretch{1}
\caption{31 $H(z)$ measurements (in units of $\rm km\,s^{-1}\,Mpc^{-1}$) obtained with the BAO method.}
\label{BAOHz}
\centering
\begin{tabular}{l p{1.8cm}<{\centering} p{1.8cm}<{\centering} p{1.8cm}<{\centering} p{1.8cm}<{\centering}}
\bottomrule[1pt]
& Redshift $z$ & $H(z)$         & $\sigma_{H(z)}$       & Reference                       \\
\bottomrule[1pt]
& 0.24      & 79.69             & 2.99                  & \citep{Gaztanaga:2008xz} \\
& 0.3       & 81.7              & 6.22                  & \citep{Oka:2013cba} \\
& 0.31      & 78.17             & 4.74                  & \citep{BOSS:2016zkm} \\
& 0.34      & 83.8              & 3.66                  & \citep{Gaztanaga:2008xz} \\
& 0.35      & 82.7              & 8.4                   & \citep{Chuang:2012qt} \\
& 0.36      & 79.93             & 3.39                  & \citep{BOSS:2016zkm} \\
& 0.38      & 81.5              & 1.9                   & \citep{BOSS:2016wmc} \\
& 0.40      & 82.04             & 2.03                  & \citep{BOSS:2016zkm} \\
& 0.43      & 86.45             & 3.68                  & \citep{Gaztanaga:2008xz} \\
& 0.44      & 82.6              & 7.8                   & \citep{Blake:2012pj} \\
& 0.44      & 84.81             & 1.83                  & \citep{BOSS:2016zkm} \\
& 0.48      & 87.79             & 2.03                  & \citep{BOSS:2016zkm} \\
& 0.51      & 90.4              & 1.9                   & \citep{BOSS:2016wmc}\\
& 0.52      & 94.35             & 2.65                  & \citep{BOSS:2016zkm} \\
& 0.56      & 93.33             & 2.32                  & \citep{BOSS:2016zkm} \\
& 0.57      & 87.6              & 7.8                   & \citep{Chuang:2013hya} \\
& 0.57      & 96.8              & 3.4                   & \citep{BOSS:2013rlg} \\
& 0.59      & 98.48             & 3.19                  & \citep{BOSS:2016zkm} \\
& 0.6       & 87.9              & 6.1                   & \citep{Blake:2012pj} \\
& 0.61      & 97.3              & 2.1                   & \citep{BOSS:2016wmc} \\
& 0.64      & 98.82             & 2.99                  & \citep{BOSS:2016zkm} \\
& 0.73      & 97.3              & 7                     & \citep{Blake:2012pj} \\
& 0.978     & 113.72            & 14.63                 & \citep{Zhao:2018gvb} \\
& 1.23      & 131.44            & 12.42                 & \citep{Zhao:2018gvb} \\
& 1.526     & 148.11            & 12.71                 & \citep{Zhao:2018gvb} \\
& 1.944     & 172.63            & 14.79                 & \citep{Zhao:2018gvb} \\
& 2.3       & 224               & 8                     & \citep{Busca:2012bu} \\
& 2.33      & 224               & 8                     & \citep{Bautista:2017zgn} \\
& 2.34      & 222               & 7                     & \citep{BOSS:2014hwf} \\
& 2.36      & 226               & 8                     & \citep{BOSS:2013igd} \\
& 2.4       & 227.8             & 5.61                  & \citep{duMasdesBourboux:2017mrl} \\
\bottomrule[1pt]
\end{tabular}
\end{table}
The Hubble parameter $H(z)$, which describes the expansion rate of the universe, can be measured in two important ways. One method is to calculate the differential ages of passively evolving galaxies (usually called cosmic chronometers), which provides the model-independent $H(z)$ measurements \citep{Jimenez:2001gg,Koksbang:2021qqc}. In the framework of general relativity, the Hubble parameter can be written in terms of the differential time evolution of the universe $\Delta t$ in a given redshift interval $\Delta z$, as
\begin{align}
\label{CC}
H(z)=-\frac{1}{1+z} \frac{\Delta z}{\Delta t}.
\end{align}
By using the CC measurements, we can obtain their redshifts and differences in age, thus achieving the estimation of $H(z)$. The CC method does not assume any fiducial cosmological model. We summarize the total 32 CC $H(z)$ measurements in Table~\ref{CCHz}. The sources of these data are quoted in the table. In should be pointed out that the CC data may not constitute a reliable source of information considering the concerns given in Ref.~\citep{Kjerrgren:2021zuo}.

{Another method is to detect the radial BAO features using the galaxy surveys and Ly-$\alpha$ forest measurements \citep{Gaztanaga:2008xz, Oka:2013cba, BOSS:2016zkm, Chuang:2012qt, BOSS:2016wmc, Blake:2012pj, Chuang:2013hya,  BOSS:2013rlg, Busca:2012bu, Zhao:2018gvb, Bautista:2017zgn, BOSS:2014hwf, BOSS:2013igd, duMasdesBourboux:2017mrl}. The BAO scale provides us with a standard ruler to measure the distances in cosmology. Note that the radial BAO measurements can only obtain the combination $H(z)r_{\rm d}$, where $r_{\rm d}$ is the sound horizon,
\begin{align}
\label{rd}
r_{\mathrm{d}}=\int_{z_{\mathrm{d}}}^{\infty} \frac{c_{\mathrm{s}}(z)}{H(z)} \mathrm{d} z,
\end{align}
evaluated at the drag epoch $z_{\rm d}$, with $c_{\mathrm{s}}$ the sound speed. In order to obtain $H(z)$, one first needs to determine the sound horizon. In this work, the fiducial value of $r_{\rm d}$ is derived from the Planck 2018 CMB observations \citep{Aghanim:2018eyx}. We compile the 31 BAO $H(z)$ data in Table~\ref{BAOHz}, which are summarized in Refs.~\citep{Mukherjee:2021ggf, Lian:2021tca}. The data set includes almost all the radial BAO data reported in various galaxy surveys. Note that some of the data points are correlated since either they belong to the same analysis or there is an overlap between galaxy samples. In this paper, we consider not only the central values and standard deviations of the BAO $H(z)$ data, but also the covariances among the data points, which are publicly available in the cited references. It can be seen that the BAO $H(z)$ data are generally more accurate than the CC $H(z)$ data. Of course, the BAO measurements also face some challenges \citep{Ellis:1987zz}, especially the environmental dependence of the BAO peak location \citep{Roukema:2014tta, Roukema:2015cwa}. We also note that there is a noticeable systematic difference between the BAO and CC $H(z)$ measurements \citep{Ding:2015vpa, Zheng:2016jlq}. Therefore, it is more reasonable to reconstruct the function of $E(z)$ using the CC and BAO $H(z)$ data, respectively. However, to tighten the constraints on the cosmic curvature, we use both CC and BAO $H(z)$ data to reconstruct $E(z)$.}

\subsection{SNe Ia}\label{sec3.2}
The data we used for reconstructing $D(z)$ and $D'(z)$ is the Pantheon compilation \citep{Pan-STARRS1:2017jku}, which contains 1048 SNe Ia covering the redshift range of $0.001 < z < 2.26$. For an SN Ia, the distance modulus $\mu$ and the luminosity distance are related by
\begin{align}
\mu(z)=5 \log \left[\frac{D_{L}(z)}{\mathrm{Mpc}}\right]+25,
\end{align}
and the observed distance modulus is
\begin{align}
\mu_{\mathrm{obs}}(z)=m_{B}(z)+\alpha \cdot X_{1}-\beta \cdot \mathcal{C}-M_{B},
\end{align}
where $m_{B}$ is the rest-frame $B$-band peak magnitude, $X_{1}$ and $\mathcal{C}$ represent the time stretch of light curve and the supernova color at maximum brightness, respectively, and $M_{B}$ is the absolute $B$-band magnitude. $\alpha$ and $\beta$ are two nuisance parameters, which could be calibrated to zero by the {\tt BEAMS} with Bias Corrections method \citep{Kessler:2016uwi}. Then the observed distance modulus can be expressed as
\begin{align}
\mu_{\mathrm{obs}}(z)=m_{B}(z) - M_{B}.
\end{align}
Once the absolute magnitude is known, the luminosity distances can be obtained.

\subsection{GW standard sirens}\label{sec3.3}
We simulate the GW standard sirens based on the spaceborne DECIGO and assume the GWs are from the binary neutron star (BNS) mergers. For the redshift distribution of BNSs, we employ the form \citep{Zhao:2010sz,Zhang:2019ple,Zhang:2019loq,Li:2019ajo,Jin:2020hmc,Jin:2021pcv,Wu:2022dgy}
\begin{align}
P(z)\propto \displaystyle{\frac{4\pi D_{C}^2(z)R(z)}{H(z)(1+z)}},
\end{align}
where $D_{C}$ is the comoving distance and $R(z)$ is the time evolution of the burst rate,
\begin{eqnarray}
R(z) =
\begin{cases}
1+2z,                                 &  z\leq 1, \\
\displaystyle{\frac{3}{4}}(5-z),      &  1 < z < 5,\\
0,                                    &  z \geq 5.
\end{cases}
\end{eqnarray}
{It should be mentioned that there are other strategies to quantify the redshift distribution of BNS \citep{Ding:2018zrk, Yang:2021qge, Belgacem:2019tbw, Ciolfi:2021gzg}.} We then calculate the fiducial value of luminosity distance in the Planck best-fit flat $\Lambda$CDM model using
\begin{align}
\label{DLDL}
D_{L}(z)=\frac{c(1+z)}{H_{0}} \int_{0}^{z} \frac{\mathrm{d} z'}{\sqrt{\Omega_{m}(1+z')^{3}+\Omega_{\Lambda}}},
\end{align}
where $H_0=67.3\ \rm km\ s^{-1}\ Mpc^{-1}$, $\Omega_{m}=0.317$, and $\Omega_{\Lambda}=0.683$. The total measurement errors of $D_L$ consist of the instrumental error, the weak lensing error, and the peculiar velocity error, i.e.,
\begin{align}
\sigma_{D_{L}}=\sqrt{(\sigma_{D_L}^{\rm inst})^2 + (\sigma_{D_{L}}^{\rm lens})^2 + (\sigma_{D_{L}}^{\rm pv})^2}.
\end{align}
For the simulation of $\sigma_{D_{L}}^{\rm inst}$, we refer the reader to Ref.~\citep{Nishizawa:2010xx}. For the error caused by the weak lensing, we adopt the form given in Ref.~\citep{Hirata:2010ba}. The error caused by the peculiar velocity of the GW source can be found in Ref.~\citep{Gordon:2007zw}. Note that we will consider the Gaussian randomness. At each redshift point $z$, the mean of luminosity distance is sampled from the normal distribution $\mathcal{N}\left(D_L(z)_{\mathrm{fid}}, \sigma_{D_L(z)}\right)$. DECIGO is expected to detect $10^5$ GW events from BNSs within the redshift range of $z\lesssim5$, as the expectation of its 1-year operation \citep{Kawamura:2020pcg}. Considering the determination of electromagnetic counterparts, we choose a normal expected scenario, i.e., $5000$ GW events with redshifts \citep{Zhang:2022lta}, as an example in this work. For a comprehensive analysis on the redshift determination of GW events from optical follow-up observations, we refer the reader to Ref.~\citep{Zhang:2022lta}. For the studies on the GW standard sirens from the coalescences of (super)massive black hole binaries based on the space-based GW observatories LISA, Taiji, and TianQin, as well as the pulsar timing arrays, see, e.g., Refs.~\cite{Wang:2019tto,Zhao:2019gyk,Wang:2021srv,TianQin:2020hid,Bian:2021ini,LISACosmologyWorkingGroup:2022jok,Wang:2022oou}.

\subsection{Future CC+BAO+RD}\label{sec3.4}
{In the future, with the advent of powerful optical and radio telescopes, we can better measure the Hubble parameter using the CC and BAO methods. In addition, the neutral hydrogen (\textsc{H\,i}) intensity mapping technique will enable us to measure the BAO signals more efficiently \citep{Bull:2014rha,Xu:2020uws,Zhang:2021yof,Wu:2021vfz,Zhang:2019ipd}. In this work, a total of 63 $H(z)$ data are considered, and we are optimistic that 200 observational data at $0<z<2.5$ will be realized in the coming decades. Following Ma \& Zhang~\citep{Ma:2010mr}, we assume that the redshift of the CC+BAO $H(z)$ subjects to a Gamma distribution and the error of $H(z)$ increases linearly with redshift. We fit the real $\sigma_{H(z)}$ data with first degree polynomial to obtain $\sigma_0(z)$, and then two lines $\sigma_{+}(z)$ and $\sigma_{-}(z)$ are selected symmetrically around it to ensure that most data points fall into the area between them. The mock $\sigma_{H(z)}$ data are generated according to the normal distribution $\mathcal{N}\left(\sigma_0(z), \ve(z)\right)$, where $\ve(z)=[\sigma_{+}(z)+\sigma_{-}(z)]/4$ is set to ensure that $\sigma_{H(z)}$ falls in the area with $2\sigma$ probability. Then the mean of $H(z)$ is sampled from $\mathcal{N}\left(H(z)_{\mathrm{fid}}, \sigma_{H(z)}\right)$. For more details, we refer the reader to Ref.~\citep{Ma:2010mr}.}

Now we turn to simulating the high-$z$ RD $H(z)$ data. In an observing time interval $\Delta t$, the shift in the spectroscopic velocity of a source $\Delta v$ can be expressed as \citep{Liske:2008ph}
\begin{align}
\Delta v=\frac{c\Delta z}{1+z}=cH_{0} \Delta t\left(1-\frac{E(z)}{1+z}\right).
\end{align}
{Loeb \citep{Loeb:1998bu} pointed out that the high-resolution spectrographs on large telescopes have the potential to measure the shifts in absorption-line spectra of distant quasi-stellar objects (QSOs). By observing the Ly-$\alpha$ absorption lines of QSOs, the E-ELT could measure the velocity shifts in the redshift range of $2<z<5$} \citep{Loeb:1998bu,Liske:2008ph,Zhang:2010im,Geng:2014hoa,Geng:2014ypa,Geng:2015ara,Geng:2015hen,Guo:2015gpa,He:2016rvp}. According to the study of Liske et al.~\citep{Liske:2008ph}, the achievable precision of $\Delta v$ can be estimated as
\begin{align}
\sigma_{\Delta v}=1.35\left(\frac{2370}{\rm{S / N}}\right)\left(\frac{N_{\mathrm{QSO}}}{30}\right)^{-1/2}\left(\frac{1+z_{\mathrm{QSO}}}{5}\right)^{q} \mathrm{cm} \mathrm{~s}^{-1},
\end{align}
where $\rm{S/N}$ is the signal-to-noise ratio of the Ly-$\alpha$ spectrum, $N_{\mathrm{QSO}}$ is the number of observed QSOs at the effective redshift $z_{\mathrm{QSO}}$, and the exponent $q$ is $-1.7$ up to $z=4$ and $-0.9$ for $z>4$. {In this work, we assume five RD measurements at effective redshifts $z=2.5$, 3.0, 3.5, 4.0, and 4.5, with $N_{\rm QSO} = 6$ in each of five redshift bins \citep{Martins:2016bbi,Alves:2019hrg}, namely a total of 30 observable quasars. In addition, we make the assumptions of $\Delta t=20$ yr and $\rm{S/N} = 3000$ for the RD measurements.} We simulate the RD $H(z)$ data in the Planck best-fit flat $\Lambda$CDM model, and the error of $H(z)$ is calculated from the precision of $\Delta v$.

\section{Results and discussions}\label{sec4}
We first need to normalize the observational Hubble parameter and luminosity distance data to get the $E(z)$ and $D(z)$ data. From Eq.~(\ref{test}), we have
\begin{align}
\label{test1}
\mathcal{O}_{K}(z)=\frac{H(z)}{\hat{H}_0} {\left[\frac{\hat{H}_0}{c(1+z)}D_{L}(z)\right]}^{\prime}-1,
\end{align}
where $\hat{H}_0$ is the normalization factor. Since the two factors can cancel out each other, whose value will not influence the null test of the cosmic curvature. In this work, we adopt $\hat{H}_0=70.0\ \rm km\ s^{-1}\ Mpc^{-1}$ to normalize the CC+BAO $H(z)$ data as the observational $E(z)$. For consistency, we use the same $\hat{H}_0$ to normalize the SNe Ia $D_L(z)$ data as the observational $D(z)$. We then adopt the GP method to reconstruct the functions of $E(z)$, $D(z)$, and $D^{\prime}(z)$, and the results are shown in Fig.~\ref{EzDzDz}. The black line is the mean of the reconstruction and the shaded grey regions are the $1\sigma$ ($68.3\%$) and $2\sigma$ ($95.4\%$) confidence level (C.L.) of the reconstruction. {One may find that the errors of the reconstructed function are significantly smaller than the errors of the data themselves. This is due to the basic assumptions that the distribution of the function at each point is Gaussian and the data points are correlated by the covariance function.} As can be seen, the error of $E(z)$ does not increase significantly with redshift due to the relatively accurate BAO data. However, the error of $D(z)$ and $D^{\prime}(z)$ becomes very large at $z>1.5$ because the data in that region are scarce and of poor quality. {In addition, all the reconstructed functions in the redshift interval $z<1$ are consistent well with a flat $\Lambda$CDM model with $\Omega_{\rm m}=0.28$, which is adopted for a comparison.}
\begin{figure*}[!htbp]
\includegraphics[scale=0.29]{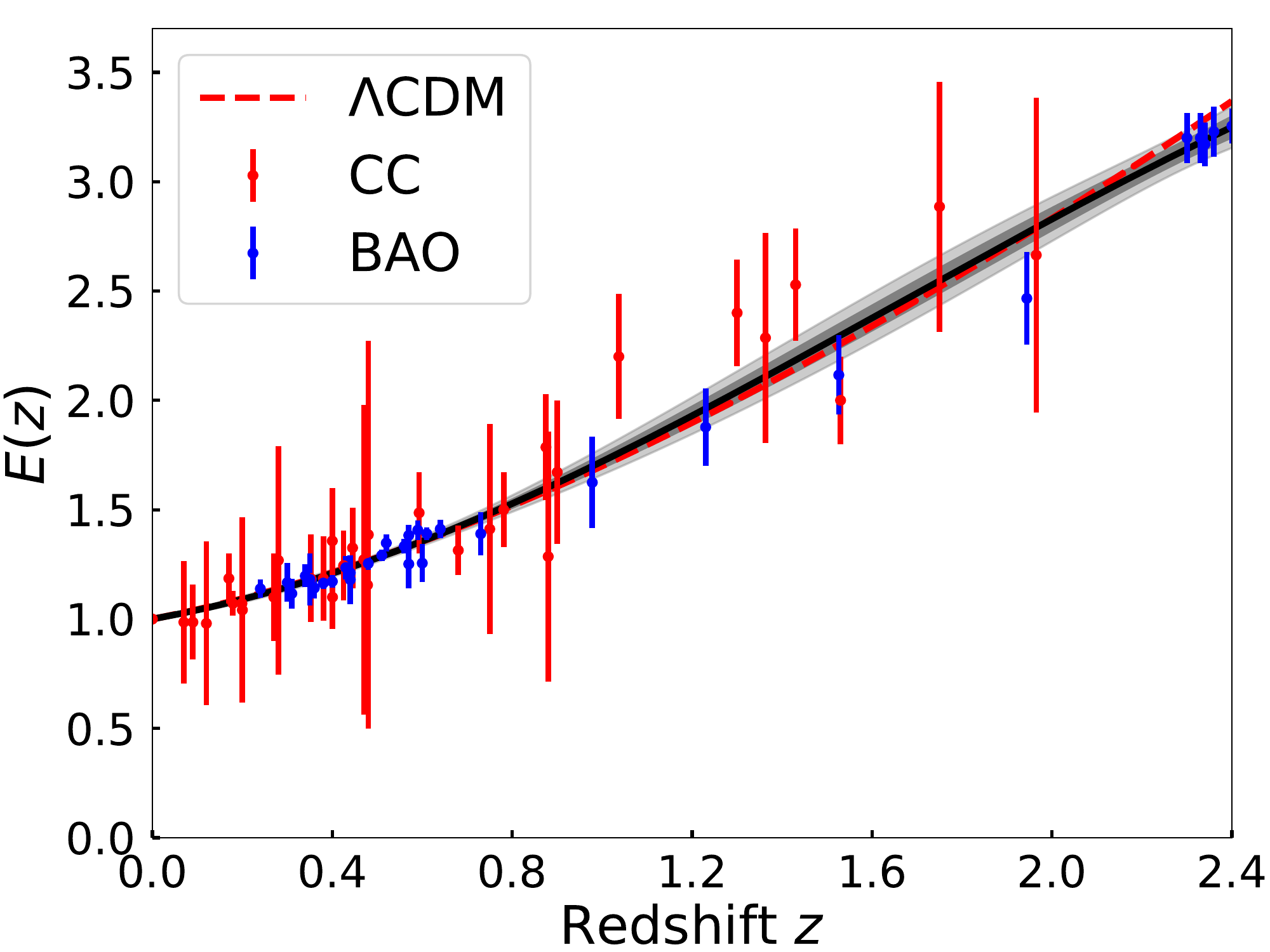}
\includegraphics[scale=0.29]{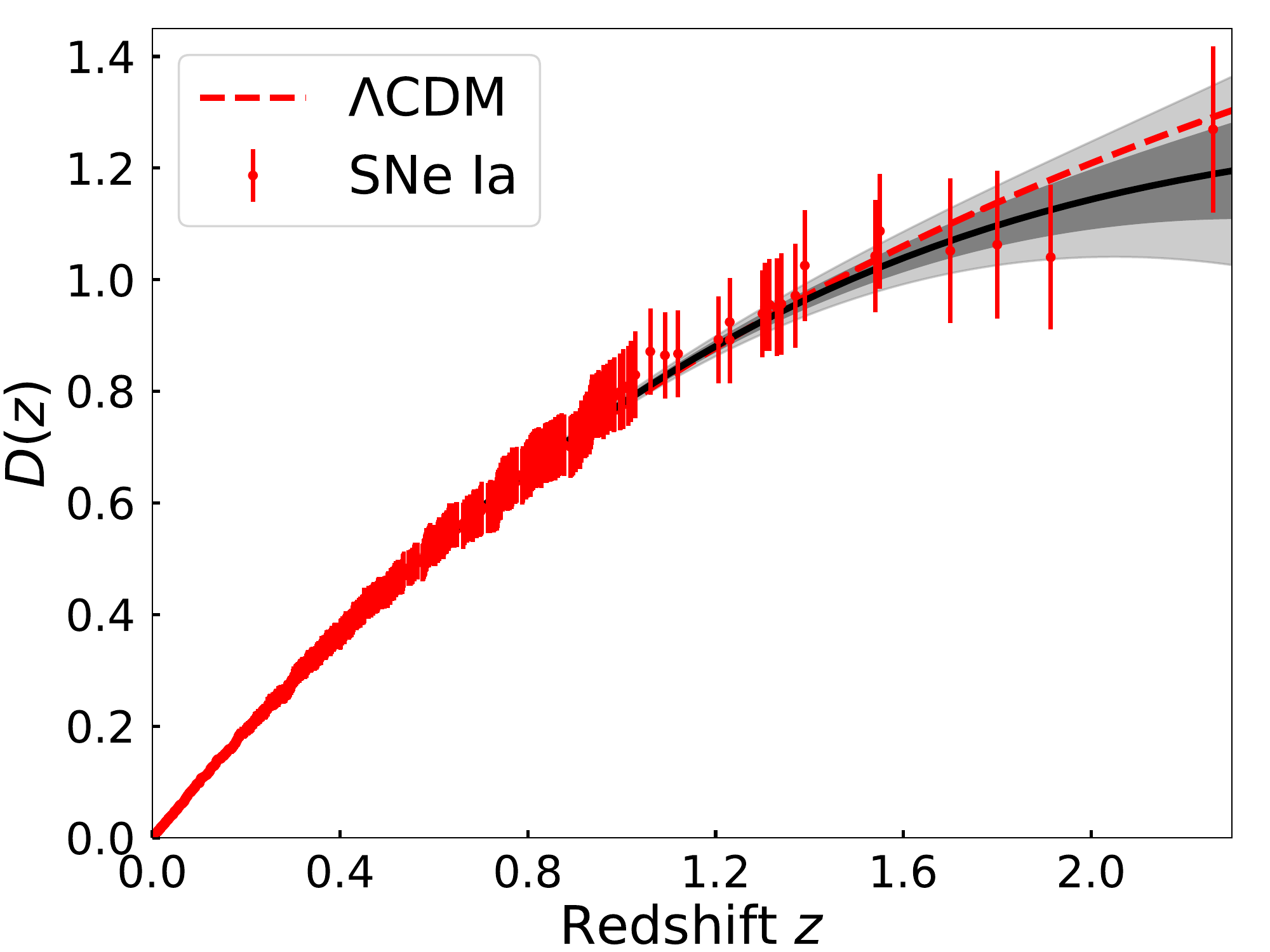}
\includegraphics[scale=0.29]{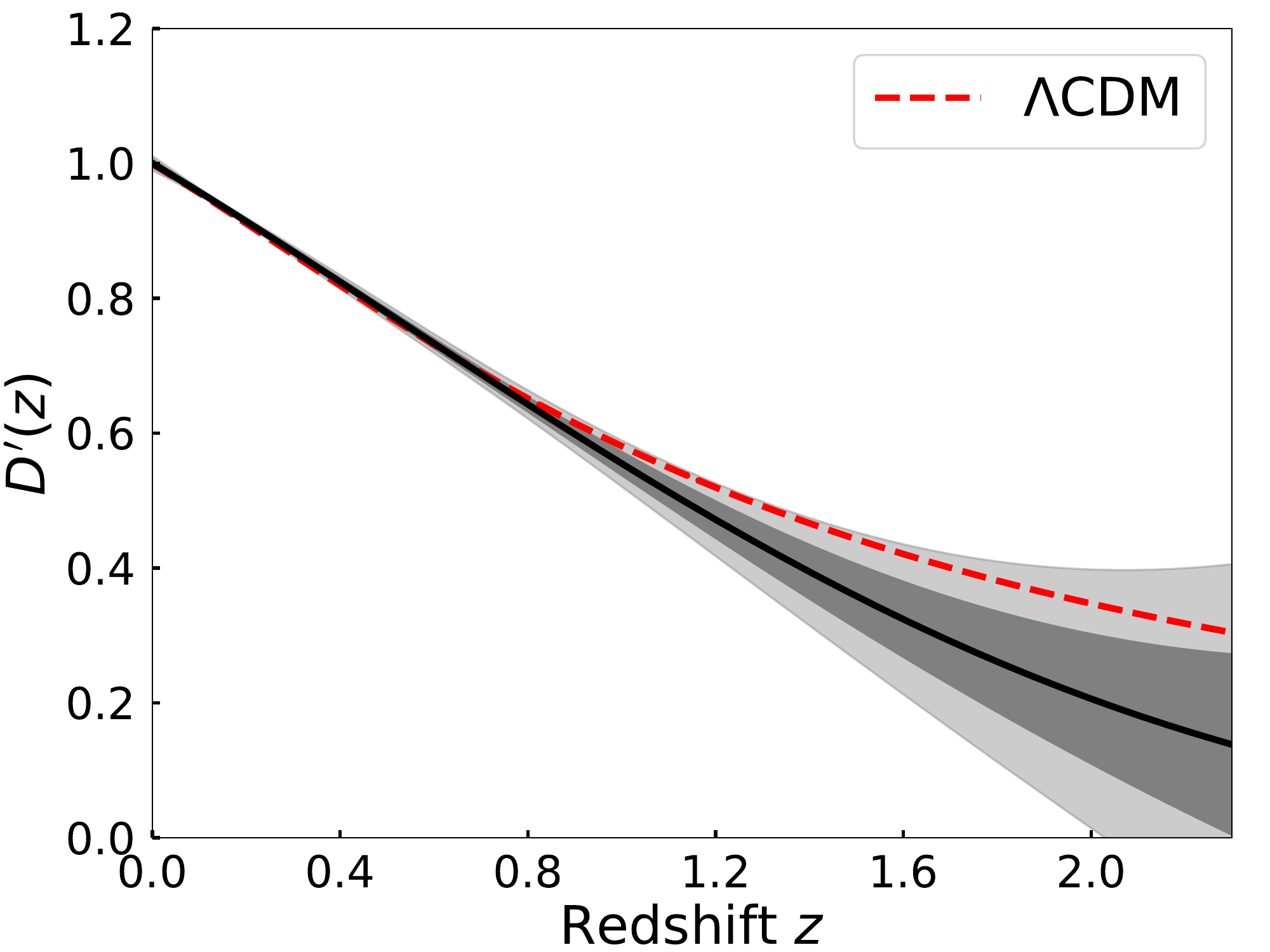}
\centering
\caption{Gaussian process reconstruction of $E(z)$ (left panel) from the CC+BAO data, $D(z)$ (middle panel), and $D'(z)$ (right panle) from the SNe Ia data. The grey shaded regions are the $1\sigma$ and $2\sigma$ C.L. of the reconstruction. The dots with error bars are the observational data. {A flat $\Lambda$CDM model (red dashed lines) is also shown for a comparison.}}
\label{EzDzDz}
\end{figure*}

\begin{figure*}[!htbp]
\includegraphics[scale=0.4]{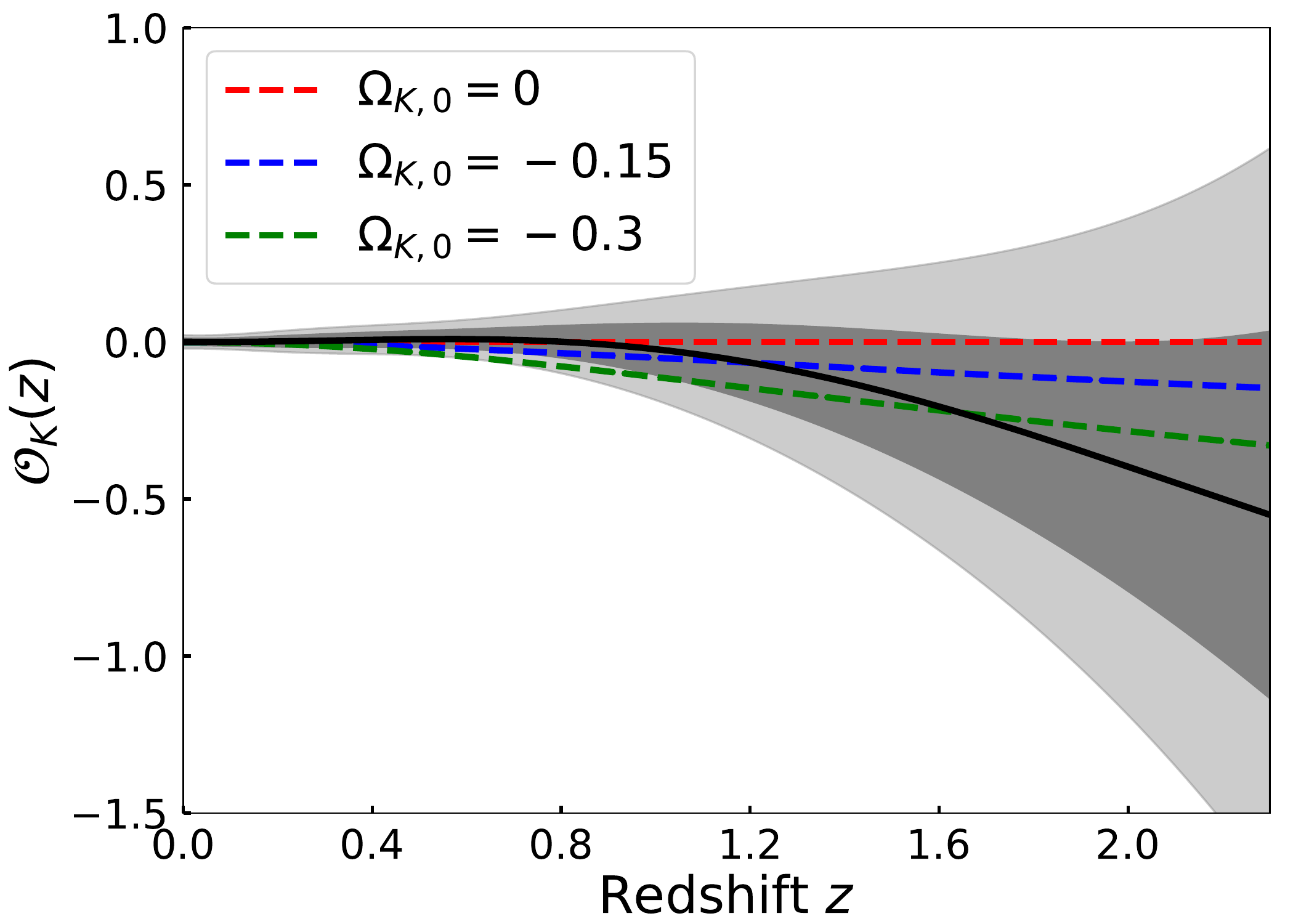}
\includegraphics[scale=0.4]{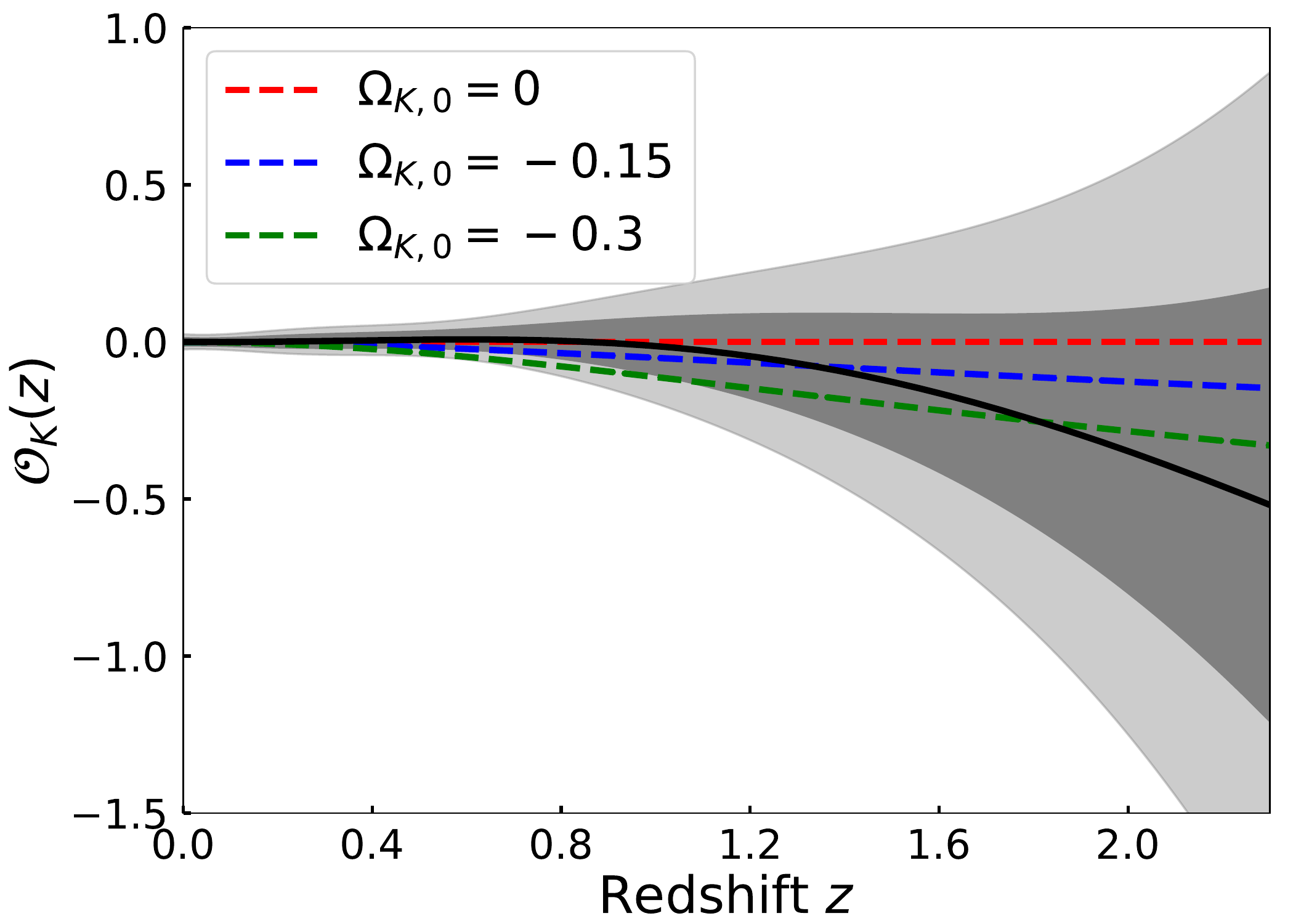}
\centering
\caption{Reconstruction of $\mathcal{O}_{K}(z)$ from the CC+BAO and SNe Ia data. The result is based on the reconstructions by GP with squared exponential function (left panel) and Matern92 covariance function (right panel). The grey shaded regions are the $1\sigma$ and $2\sigma$ C.L. of the reconstruction. The red dashed line corresponds to a flat universe with $\Omega_{K,0}=0$. {The blue and green dashed lines correspond to the cases of a $\Lambda$CDM model with $\Omega_{K,0}=-0.15$ and $\Omega_{K,0}=-0.3$, respectively.}}
\label{result}
\end{figure*}

With the GP reconstructions, we can then carry out the null test of the cosmic curvature. In Ref.~\citep{Cai:2015pia}, the authors applied the Monte Carlo sampling to determine $\mathcal{O}_{K}(z)$. Different from them, we use the error propagation formula to calculate the error of $\mathcal{O}_{K}(z)$ at each point $z$. The result is shown in the left panel of Fig.~\ref{result}. The red dashed line refers to the spatially flat universe with $\Omega_{K,0}=0$. It can be seen that the result is consistent with the flat universe within the domain $0<z<2.3$, falling within the $1\sigma$ C.L. We note that the mean of $\mathcal{O}_{K}(z)$ is very close to zero in the redshift interval $0<z<1$, however, it becomes more and more negative at $z>1$. In addition, the universe with $\Omega_{K,0}=0$ almost falls out of the $1\sigma$ C.L. at $z>1.5$. This indicates that there is still a possibility for a spatially closed universe. {Notably, the reconstruction at $z>1.5$ prefers a closed universe over an open one. We also plot the $\Lambda$CDM model with negative curvature in Fig.~\ref{result}. As can be seen, a universe with $\Omega_{K,0}=-0.15$ can fall within the $1\sigma$ C.L. and a universe with $\Omega_{K,0}=-0.3$ can fall within the $2\sigma$ C.L. Note that the curves shown in Fig.~\ref{result} are plotted by assuming a $\Lambda$CDM model (with $\Omega_{\rm m}=0.28$) which is adopted here for a comparison.}

In the GP analysis above, we only consider the squared exponential covariance function. In fact, the choice of covariance function will influence the result. To illustrate the effect, we take the Matern92 covariance function as an example, which is given by
\begin{align}
\begin{aligned}
k(z, \tilde{z})=& \sigma_{f}^{2} \exp \left(-\frac{3|z-\tilde{z}|}{\ell}\right)\left(1+\frac{3|z-\tilde{z}|}{\ell}+\frac{27(z-\tilde{z})^{2}}{7 \ell^{2}}\right.\\
&\left.+\frac{18|z-\tilde{z}|^{3}}{7 \ell^{3}}+\frac{27(z-\tilde{z})^{4}}{35 \ell^{4}}\right).
\end{aligned}
\end{align}
Following the process described above, we perform the null test again, and the updated result is shown in the right panel of Fig.~\ref{result}. The result is still consistent with the flat universe, falling within the $1\sigma$ C.L. At $z>1$, the mean of $\mathcal{O}_{K}(z)$ also deviates from zero and becomes negative. The difference is that the error of $\mathcal{O}_{K}(z)$ using the Matern92 covariance function is slightly larger than that using the squared exponential covariance function. In general, the difference does exist, but it is not significant enough to change our conclusions. In the following, we adopt only the GP with squared exponential covariance to complete our analysis.

The large error of $\mathcal{O}_{K}(z)$ leaves the big window open for a possible non-flat universe. Therefore, it is necessary to forge new cosmological probes to precisely measure the Hubble parameter and luminosity distance, thus tightening the constraints on the cosmic curvature. In the next decades, the GW standard siren and RD method will be greatly developed. We can adopt them to measure $D_L(z)$ and $H(z)$. {Then, it is worth further studying the issue of combining the GW and RD observations with the traditional CC and BAO measurements to test the spatial flatness of the universe.} We normalize the simulated CC+BAO+RD $H(z)$ data as the $E(z)$ data, and normalize the GW $D_L(z)$ data as the $D(z)$ data. Note that the error of RD $H(z)$ is relatively small and decreases with redshift, which is very helpful for us to reconstruct the cosmological function. Having obtained the data sets of $E(z)$ and $D(z)$, we adopt GP to reconstruct $E(z)$, $D(z)$, and $D'(z)$, and the results are shown in Fig.~\ref{EzDzDzGW}. As can be seen, the error of $E(z)$ does not increase significantly even up to the redshift of $4.5$. In addition, the error of $D'(z)$ reconstructed from the GW data is obviously smaller than that from the SNe Ia data. {Here, we wish to note that although the RD method is promising in measuring the Hubble parameter, the actual measurement will be fairly challenging even with the powerful facilities such as the E-ELT.}

\begin{figure*}[!htbp]
\includegraphics[scale=0.29]{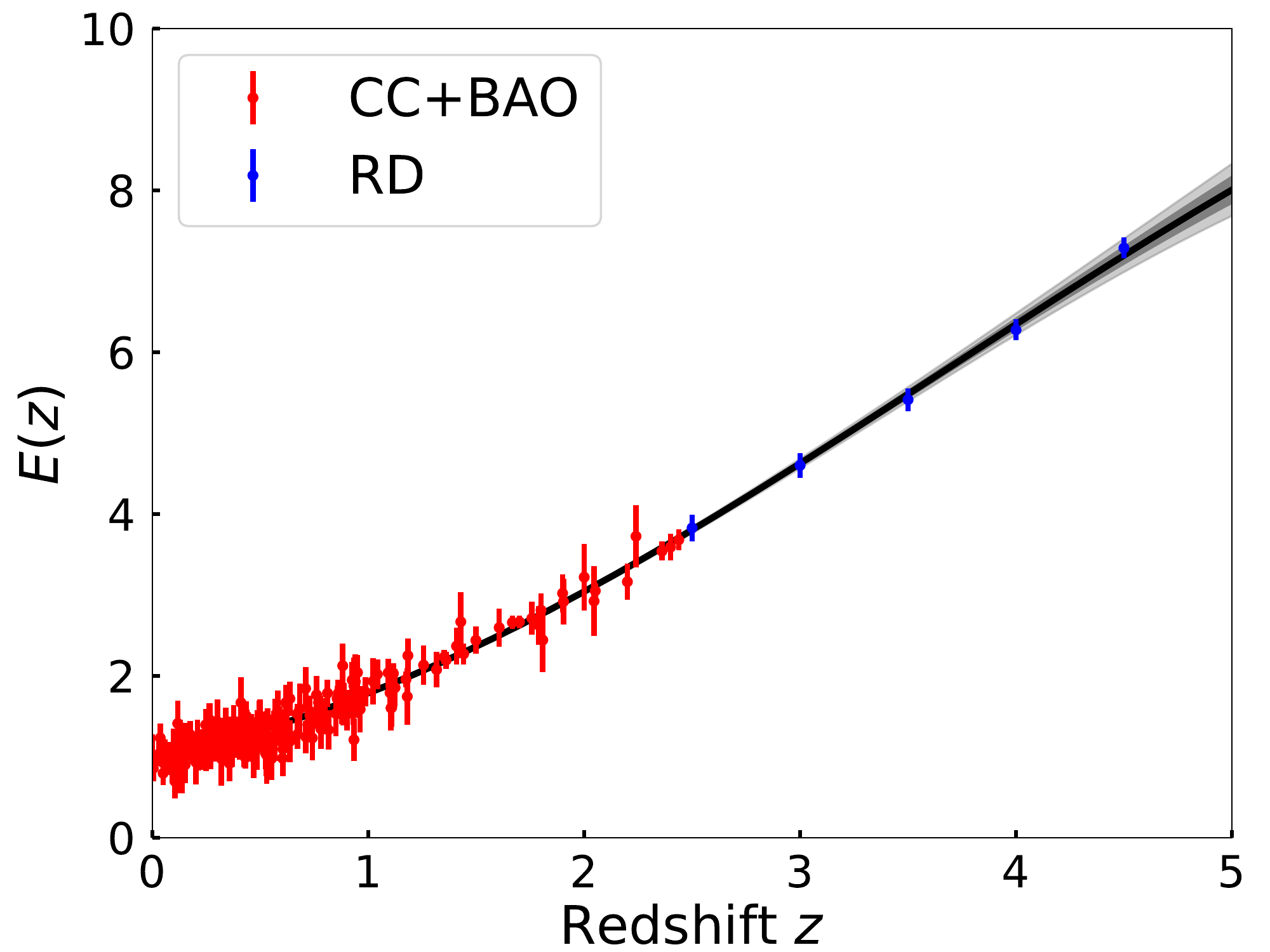}
\includegraphics[scale=0.29]{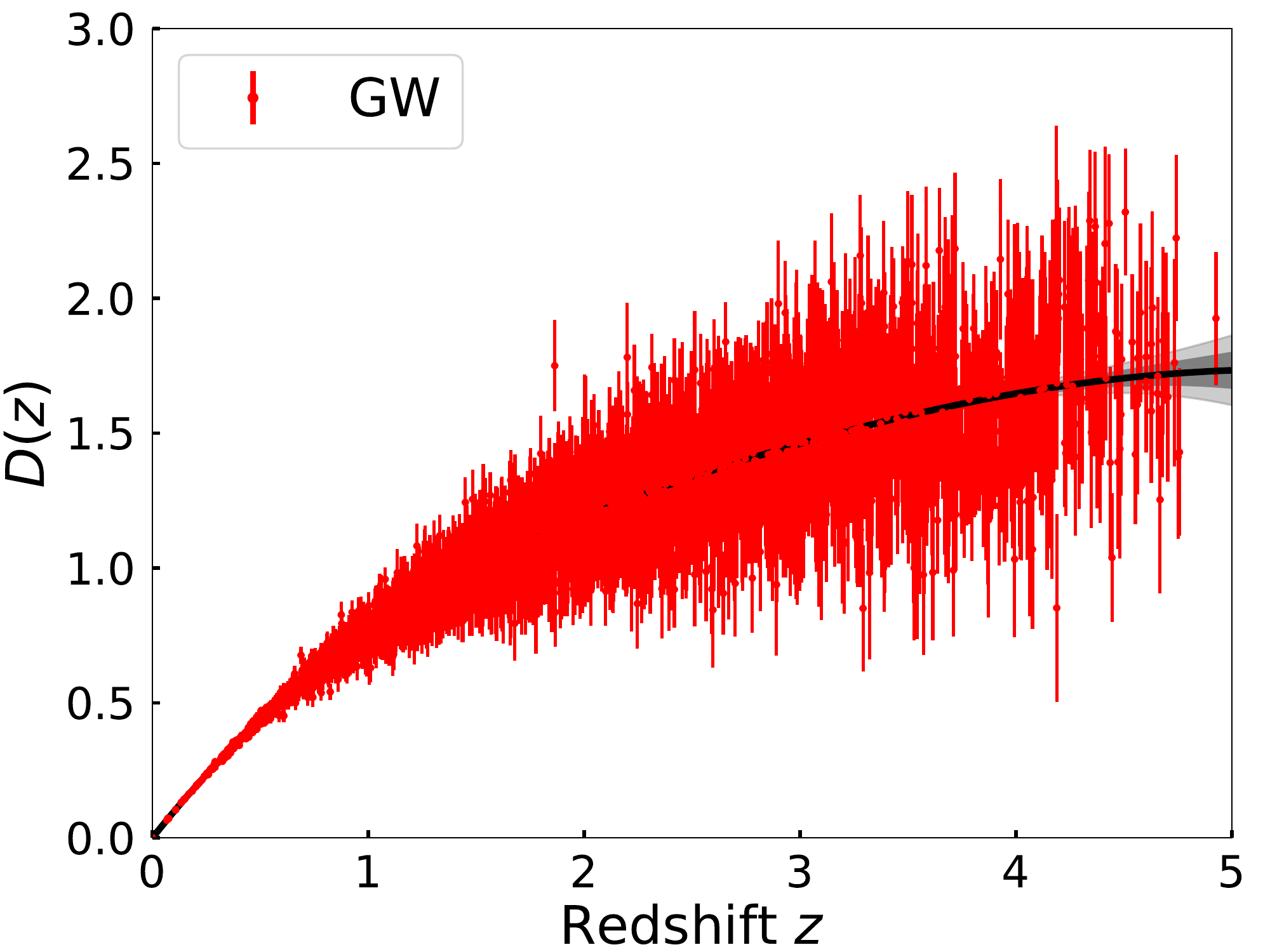}
\includegraphics[scale=0.29]{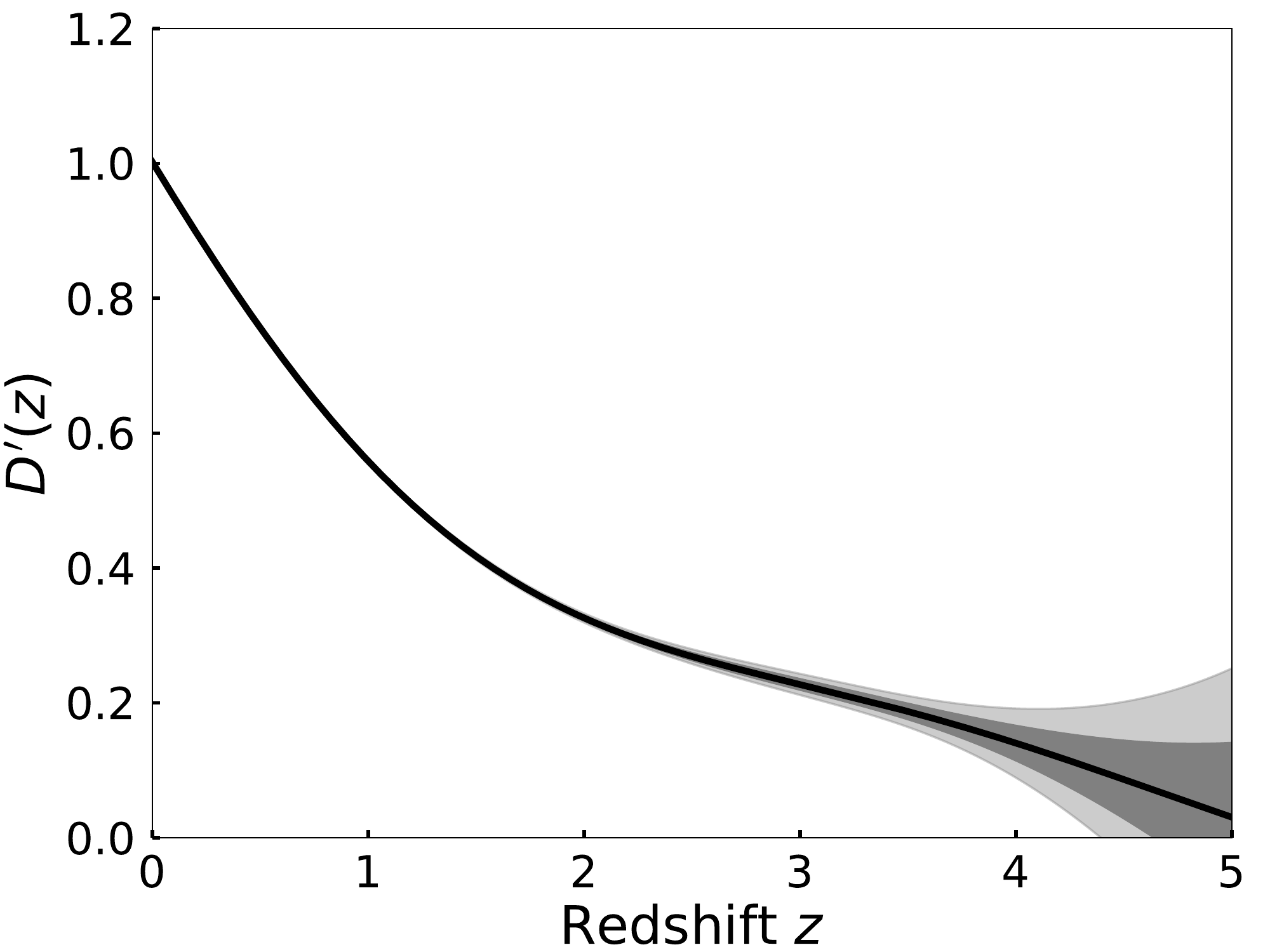}
\centering
\caption{Gaussian process reconstruction of $E(z)$ (left panel) from the simulated CC+BAO+RD $H(z)$ data, $D(z)$ (middle panel), and $D'(z)$ (right panle) from the simulated GW $D_L(z)$ data. The grey shaded regions are the $1\sigma$ and $2\sigma$ C.L. of the reconstruction. The dots with error bars are the simulated data.}
\label{EzDzDzGW}
\end{figure*}
\begin{figure*}[!htbp]
\includegraphics[scale=0.4]{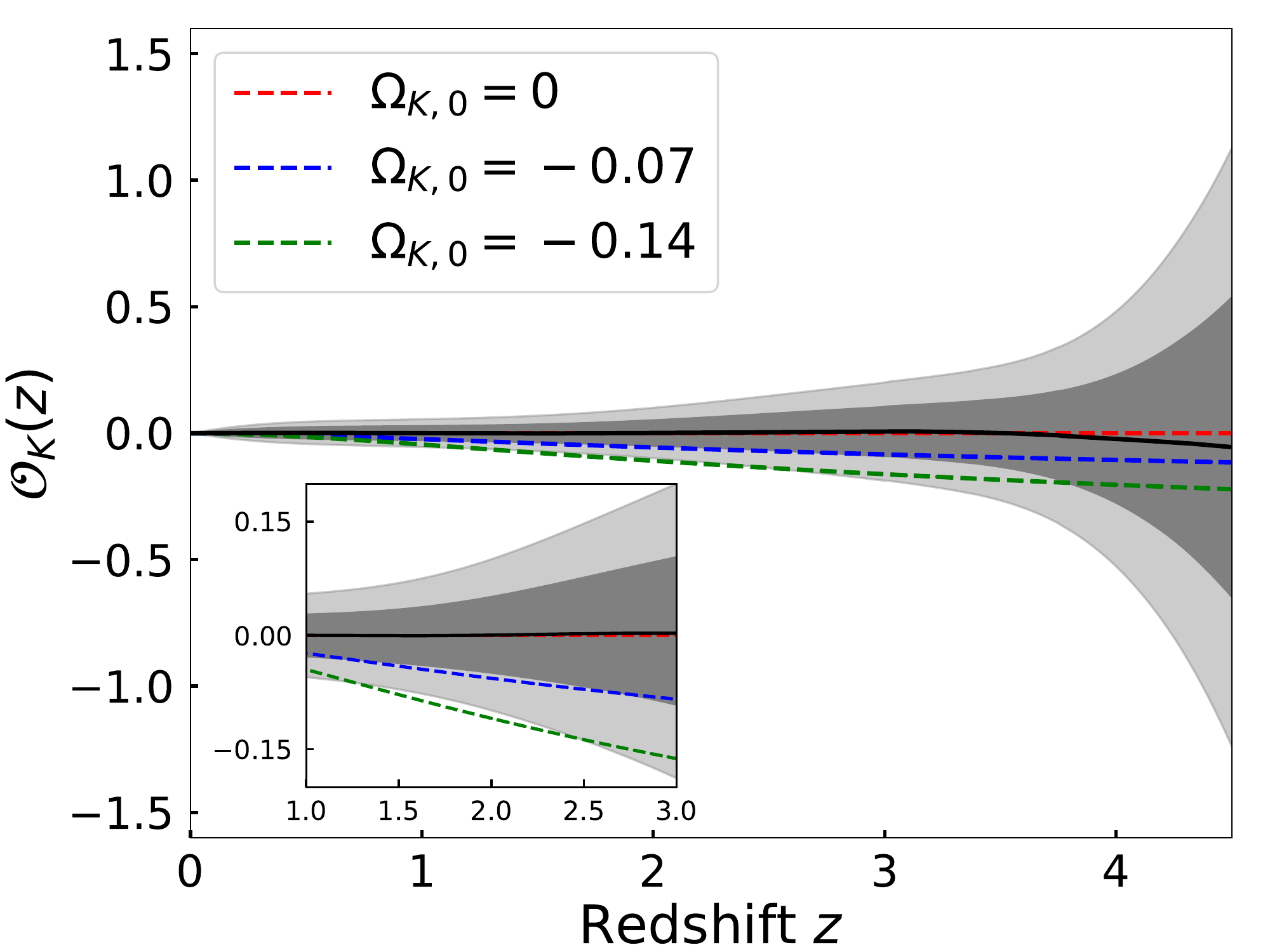}
\includegraphics[scale=0.4]{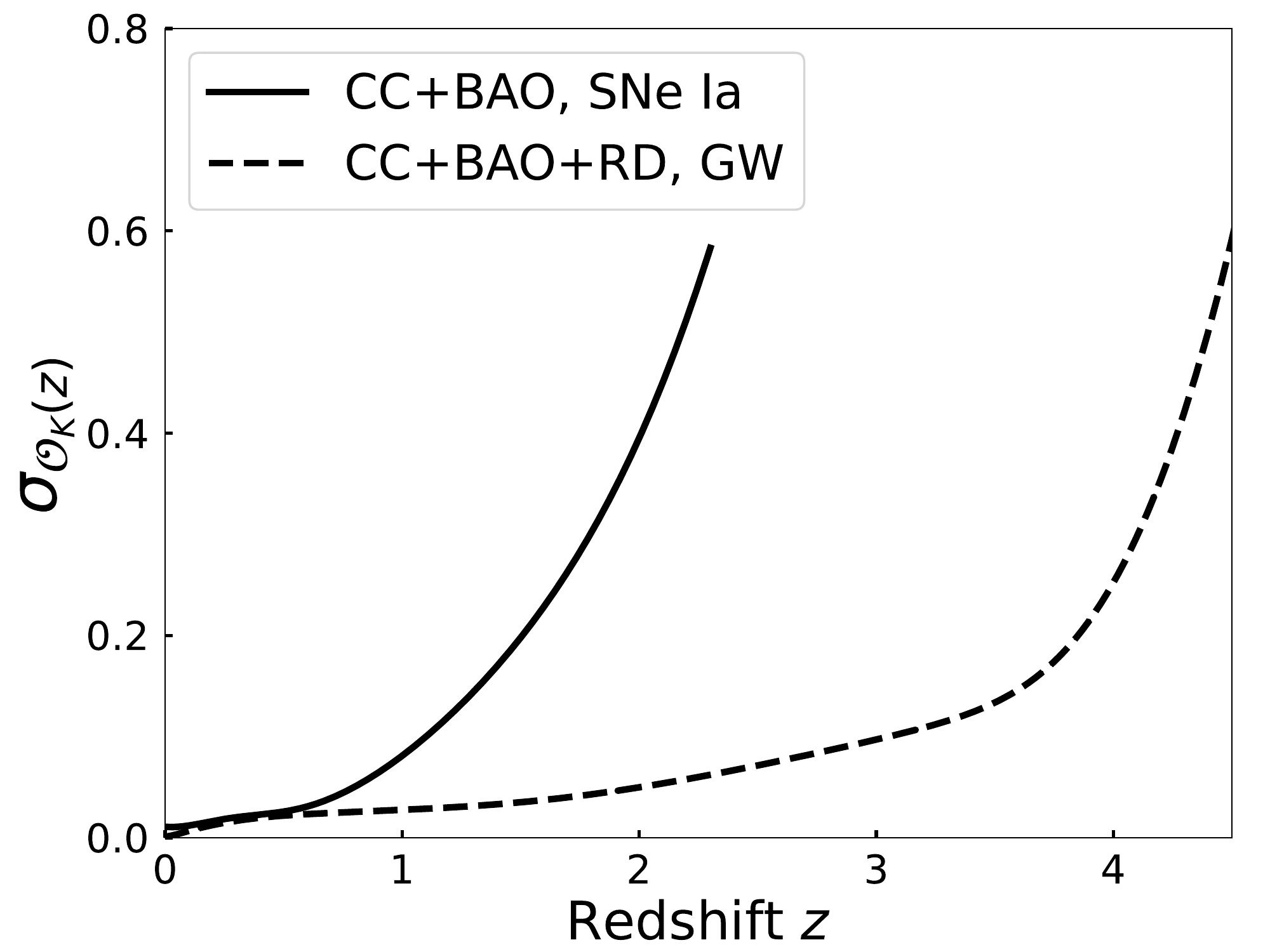}
\centering
\caption{Reconstruction of $\mathcal{O}_{K}(z)$ from the simulated GW and CC+BAO+RD data (left panel). The grey shaded regions are the $1\sigma$ and $2\sigma$ C.L. of the reconstruction. The red, blue, and green dashed lines correspond to the universe with $\Omega_{K,0}=0$, $\Omega_{K,0}=-0.07$, and $\Omega_{K,0}=-0.14$, respectively. The right panel shows the $1\sigma$ error of the reconstructed $\mathcal{O}_{K}(z)$ from the observational \{CC+BAO, SNe Ia\} data and simulated \{CC+BAO+RD, GW\} data, respectively.}
\label{resultGW}
\end{figure*}

With the reconstructions of $E(z)$ and $D'(z)$, we use the error propagation formula to determine $\mathcal{O}_{K}(z)$, and the result is shown in the left panel of Fig.~\ref{resultGW}. We see that the mean of $\mathcal{O}_{K}(z)$ is very close to zero in the redshift interval $0<z<4.5$. Some weak deviations are mainly due to the consideration of Gaussian randomness in the simulations. The result strongly favors a flat universe, which is consistent with the assumed flat $\Lambda$CDM model. Note that we only consider $\mathcal{O}_{K}(z)$ at $0<z<4.5$. Even though we have reconstructed the function of $E(z)$ at $0<z<5$, the part of $z>4.5$ is extrapolated, whose accuracy cannot be guaranteed. On the other hand, the GW $D(z)$ data at $z>4.5$ are scarce and of poor quality, so the reconstruction is not convincing. We also plot the $\Lambda$CDM model with negative curvature in the left panel of Fig.~\ref{resultGW}. As can be seen, the reconstruction at $1.5<z<2.5$ can rule out the universe with $\Omega_{K,0}=-0.07$ at $1\sigma$ C.L. and the universe with $\Omega_{K,0}=-0.14$ at $2\sigma$ C.L. Similarly, we can rule out the universe with positive curvature in this way. We have tested that if the number of CC+BAO $H(z)$ data reaches 300, the reconstruction at $1.5<z<2.5$ can rule out the universe with $\Omega_{K,0}=-0.05$ at $1\sigma$ C.L. and the universe with $\Omega_{K,0}=-0.09$ at $2\sigma$ C.L. We compare the $1\sigma$ error of $\mathcal{O}_{K}(z)$ derived from the different data in the right panel of Fig.~\ref{resultGW}. It can be seen that the error of $\mathcal{O}_{K}(z)$ derived from the future \{CC+BAO+RD, GW\} data is significantly smaller than that from the current \{CC+BAO, SNe Ia\} data. Concretely, for example, the error provided by \{CC+BAO+RD, GW\} is less than that given by \{CC+BAO, SNe Ia\} at $z=1$ by 66.6\% (and at $z=2$ by $87.3\%$). Moreover, the error of $\mathcal{O}_{K}(z)$ derived from \{CC+BAO, SNe Ia\} grows rapidly at $z>1.5$, while the error of $\mathcal{O}_{K}(z)$ from \{CC+BAO+RD, GW\} does not grow rapidly until $z\sim4$. All these analyses indicate that with the synergy of multiple high-quality observations in the future, we can better determine the spatial topology of the universe.

{In this work, we simulated only five RD measurements at high redshifts by observing the Ly-$\alpha$ absorption lines of QSOs. It should be pointed out that the SKA Phase I can measure the RD at $0<z<0.3$ by observing the \textsc{H\,i} emission lines of galaxies \citep{Weltman:2018zrl, Liu:2019asq, Qi:2021iic}. However, due to the excellent performance of CC+BAO in reconstructing $E(z)$, we did not consider this case. In addition, we did not consider the future observations of supernovae, because the GW data can reconstruct $D(z)$ and $D'(z)$ very well, as shown in Fig.~\ref{EzDzDzGW}. We note that the GW standard siren method also has the potential to measure the Hubble parameter \citep{Nishizawa:2010xx}. In principle, the observations of luminosity distances to GW sources across the sky should not be directional. However, mainly due to the local motion of the observer, there are tiny anisotropies in the luminosity distance, which enables us to measure $H(z)$. In this work, we considered a conservative scenario, i.e., $5000$ GW events with determined redshifts. In very optimistic scenarios, DECIGO is expected to detect $10^5\sim10^6$ GW events with determined redshifts. If that can be done, we can reconstruct the functions of $D(z)$ and $D'(z)$ with breathtaking precision and measure the Hubble parameter at $0<z\lesssim3$ with a few percent accuracy \citep{Nishizawa:2010xx}. Then we can test the spatial flatness of the universe in a {cosmological} model-independent way using only the GW data. We plan to explore this possibility in a future work.}

\section{Conclusions}\label{sec5}
In this paper, we adopt a {cosmological model-independent method} to test whether the cosmic curvature $\Omega_{K,0}$ deviates from zero. We use the Gaussian process method to reconstruct the reduced Hubble parameter $E(z)$ and the distance-redshift relation $[D(z),D'(z)]$, independently. {In the reconstruction, we do not assume any specific cosmological model.} By combining the reconstructions of $E(z)$ and $D'(z)$, we can determine $\mathcal{O}_{K}(z)$, which is zero at any redshift for a spatially flat universe with $\Omega_{K,0}=0$. Thus, we can carry out the null test of $\Omega_{K,0}$. We adopt the latest CC Hubble data, radial BAO Hubble data, and Pantheon SNe Ia data to implement our analysis.

Our result is consistent with a universe with $\Omega_{K,0}=0$ within the domain of reconstruction $0<z<2.3$, falling within the $1\sigma$ confidence level. We stress that the reconstruction favors a flat universe at $0<z<1$, however, it tends to favor a closed universe at $z>1$. In this sense, there is still a possibility for a closed universe. The error of the reconstructed function grows rapidly at $z>1.5$ due to the poor-quality observational data in that region, so it is necessary to forge new cosmological probes to precisely measure the luminosity distance and Hubble parameter. The GW standard siren and redshift drift observations that can be used to measure $D_L(z)$ and $H(z)$ will be greatly developed in the next decades. We simulated the GW standard siren and RD data based on the hypothetical observations of the upcoming DECIGO and E-ELT, respectively. {The traditional methods for measuring the Hubble parameter are also promising, and we simulated the CC+BAO $H(z)$ data for the next decades. Combining these mock data, we performed the flatness test of the universe.} We find that with the synergy of multiple high-quality observations in the future, we can tightly constrain the spatial geometry of the universe or exclude the flat universe with $\Omega_{K,0}=0$.

\begin{acknowledgments}
We thank Purba Mukherjee, Bo-Yang Zhang, Bo Wang, Tian-Nuo Li, Shang-Jie Jin, Ling-Feng Wang, and Ze-Wei Zhao for fruitful discussions. We sincerely thank Purba Mukherjee for providing us with the BAO data.
This work was supported by
the National SKA Program of China (Grants Nos. 2022SKA0110200 and 2022SKA0110203) and
the National Natural Science Foundation of China (Grants Nos. 11975072, 11835009, and 11875102).
\end{acknowledgments}

\bibliography{Null_test}

\end{document}